\documentclass{article}
\usepackage{locata,amsmath,graphicx,url,times}
\usepackage{amssymb,amsmath,bm}
\usepackage{algorithm}
\usepackage{algorithmic}
\usepackage{amsmath}

\DeclareMathOperator*{\argmin}{argmin} 

\title{Intensity Particle Flow SMC-PHD Filter For Audio Speaker Tracking}

\twoauthors
  {Yang Liu, Wenwu Wang}
  {University of Surrey\\
  Centre for Vision, Speech and Signal Processing\\
  Guildford, GU2 7XH, U.K\\
  $[$yangliu, w.wang$]$@surrey.ac.uk}
  {Volkan Kilic}
  {Izmir Katip Celebi University\\
  Department of Electrical and Electronics Engineering\\
  35620 Cigli-Izmir, Turkey\\
  volkan.kilic@ikc.edu.tr}

\begin{document}

\ninept
\maketitle

\begin{sloppy}

\begin{abstract}
Non-zero diffusion particle flow Sequential Monte Carlo probability hypothesis density (NPF-SMC-PHD) filtering has been recently introduced for multi-speaker tracking. However, the NPF does not consider the missing detection which plays a key role in estimation of the number of speakers with their states. To address this limitation, we propose to use intensity particle flow (IPF) in NPF-SMC-PHD filter. The proposed method, IPF-SMC-PHD, considers the clutter intensity and detection probability while no data association algorithms are used for the calculation of particle flow. Experiments on the LOCATA (acoustic source Localization and Tracking) dataset with the sequences of task 4 show that our proposed IPF-SMC-PHD filter improves the tracking performance in terms of estimation accuracy as compared to its baseline counterparts.

\end{abstract}


\section{Introduction}

The problem of acoustic source localization and tracking in an enclosed space has attracted an increased amount of attention in the last decade due to its potential applications such as advanced computer interfaces \cite{yeo2015hand}, hearing aids \cite{luts2010multicenter} and speech recognition \cite{yang2017audiovisual}. To address this problem, several methods, such as direction of arrival (DOA) \cite{talantzis2005estimation}, generalized cross-correlation (GCC) phase transform (PHAT) \cite{qin2008subsample}, steered response power (SRP) PHAT, beam steering \cite{johansson2005robust}, and time delay of arrival (TDOA) estimates \cite{ma2006tracking}, have been proposed. The trajectories of the speakers can be extracted using estimated positions by aforementioned methods. However, these trajectories may involve the random errors, false returns from background clutters, and detection loss \cite{kilicc2015audio}. To overcome these issues, filters are used to smooth the estimated trajectories. Representative filters include Kalman \cite{haykin2001kalman} and particle \cite{gustafsson2002particle} filters employed in tracking of a single moving sound source.

To track multiple moving sources, the unknown and variable number of sources need to be handled for reliable tracking. Therefore, PHD filter \cite{mahler2000theoretical} and its extension such as cardinalized PHD filter \cite{mahler2007phd} are elegant solutions for multiple source tracking. The Gaussian mixture (GM) \cite{vo2006} and sequential Monte Carlo (SMC) \cite{vo2005sequential} are the implementations to obtain practical solutions of the PHD filter. \cite{vo2005sequential}. However, it suffers from the weight degeneracy problem \cite{daum2007nonlinear}. To address this problem, particle flow is proposed \cite{daum2007nonlinear}, which migrates particles from the prior distribution to the posterior distribution based on a homotopy function defined for particle flow. In the literature, particle flow is categorized into five main classes: incompressible particle flow \cite{daum2007nonlinear}, zero diffusion particle flow (ZPF) \cite{bunch2016approximations}, Coulomb law  particle flow \cite{Daum2011b}, zero-curvature particle flow \cite{daum2013zero} and non-zero diffusion particle flow (NPF) \cite{Daum2013c}. Recently, ZPF-SMC-PHD and NPF-SMC-PHD filters are used to track multi-speakers based on the audio-visual information \cite{yang2017audiovisual,yang2018particle}.

For acoustic source tracking, the filters are mostly conducted with simulated data \cite{lollmann2018locata}. For the objective benchmarking of state-of-the-art algorithms on real-world data, the LOCATA dataset under the IEEE AASP Challenge is released \cite{lollmannlocata}. The dataset comprises six tasks ranging from the tracking of a single static sound source to the tracking of multiple moving speakers. It contains real-world audio recordings obtained by DICIT array, Eigenmike array, Robot head and Hearing aids in an enclosed acoustic environment. The sound sources are represented by moving human talkers or static loudspeakers.

In this paper, we propose a new algorithm for multi-speaker tracking, namely IPF-SMC-PHD filter for the task 4 of the LOCATA dataset. This task covers the multiple moving talkers using a static microphone array. The proposed method considers the clutter intensity and detection probability while no data association algorithms are used for the calculation of particle flow. The DOA lines are employed as the measurements of the IPF-SMC-PHD filter for multi-speaker tracking under challenging conditions such as occlusion. The speaker identity is estimated using the target position under the assumption that it is not changed abruptly in subsequent frames. Our methods are tested on all sub-arrays of task 4.

The reminder of this paper is organized as follows: the next section introduces the NPF-SMC-PHD filter. Section III describes our proposed IPF-SMC-PHD filtering algorithm. In Section IV, experiments on the LOCATA dataset are presented to show the performance of the proposed IPF-SMC-PHD algorithm as compared with the baseline algorithms.

\section{PROBLEM STATEMENT AND BACKGROUND}
\label{sec:format}

This section describes our problem formulation and the NPF-SMC-PHD filter. For the LOCATA challenge, we assume that the target dynamics and measurements are described as:
\begin{equation}
\label{trans}
{ \tilde { { \bm{m} } }  }_{ k }={ \bm{f} }_{ \tilde { \bm{m} }  }\left( { \tilde {  \bm{m} }  }_{ k-1 },{ \bm{ \tau  } }_{ k } \right),
\end{equation}
\begin{equation}
{ \bm{z} }_{ k }=\bm{f_z}\left( { \tilde{\bm{m}} }_{ k } ,{  \bm{\varsigma  }  }_{ k }\right)
\end{equation} 
where ${ \tilde{\bm{m}} }_{ k } \in \mathbb{R}^{4}$ is the target state vector in time $k$, defined as ${ \tilde{\bm{m}} }_{ k } = [x_k ,y_k , \dot { x }_k, \dot { y }_k ]^T$, which consists of the source azimuth $x$, elevation $y$ and the angular velocity $(\dot { x }_k ,\dot { y }_k)$. $\tilde { \quad  }  $ is used to distinguish the target state from the particle state used later. Let $\bm{Z}_k$ denote the set of DOA calculated by Multiple Signal Classification in time $k$. $\bm{Z}_k = \{\bm{z}^1_k,\bm{z}^2_k,...,\bm{z}^{R_k}_k\}$ where $R_k$ is the number of measurements at time $k$. The measurement ${ \bm{z} }^r_{ k }$ is a noisy version of the position $(x_k ,y_k)$, where $r$ is the index of the measurement. ${ \bm{\tau }  }_{ k }$ and ${ \bm{\varsigma }  }_{ k }$ are system excitation and measurement noise terms, respectively. $\bm{f}_{\tilde{\bm{m}}}$ is a transition model and $\bm{f_z}$ is a measurement model. 

In the NPF-SMC-PHD filter \cite{yang2018particle}, target PHD is approximated by $N_{k-1}$ survival particles $\{\bm{m}^{i}_{k-1}\}^{N_{k-1}}_{i=1}$ and their weights $\{\omega^{i}_{k-1}\}^{N_{k-1}}_{i=1}$ at time $k-1$. In the prediction step, the particle set is obtained by the proposal distribution $q_k$,

\begin{equation}
\label{SMCprediction1}
\bm{m}^i_{k|k-1} \sim q_k(\cdot |\bm{m}^i_{k-1},\bm{Z}_k)
\end{equation}
The proposal weights are
\begin{equation}
\label{SMCprediction3}
\omega^i_{k|k-1} = q_s\omega^i_{k-1}
\end{equation}
where $q_s$ is the surviving possibility. $N_{B}$ born particles are sampled by the importance function $p_k$,
\begin{equation}
\label{SMCbirth}
    \bm{m}^i_{k|k-1} \sim p_k(\cdot|\bm{Z}_k)
\end{equation}
The born particle weights are 
\begin{equation}
\label{SMCbirthweight}
\begin{matrix} \omega^i_{k|k-1} = \frac { \gamma _k(\bm{m}^i_{k|k-1})}{ N_{B}p_k(\bm{m}^i_{k|k-1}|\bm{Z}_k) } &  
i=N_{k-1} + 1,...,N_{k-1} +N_{B}  \end{matrix} 
\end{equation}
where $\gamma _k(\cdot)$ is the born possibility. $N_{k-1}$ is the number of surviving particles at time $k-1$. 

After predicting particles, a particle flow mitigates particle states via the Ito stochastic differential equation \cite{daum2010exact}:
\begin{equation}
\label{updatem2}
    \triangle \bm{m}^i_{k|k-1} = \bm{f}^i_k(\bm{m}^i_{k|k-1},\lambda)\triangle{\lambda}+\upsilon^i_k \bm{w}^i_k
\end{equation}
where $\bm{f}^i_k\in \mathbb{R}^{4}$ is the particle flow vector which moves the particle $\bm{m}^i_{k|k-1}$ with the distance $\triangle \bm{m}^i_{k|k-1}$ at $\lambda$. $\bm{w}^i_k \in \mathbb{R}^{4}$ is the Wiener process with the diffusion coefficient $\upsilon^i_k$, $\lambda$, called the synthetic time, takes values from $\left[0, \triangle \lambda, 2\triangle \lambda,\cdots , N_{ \lambda  } \triangle \lambda \right]$ and $ N_{ \lambda  } \triangle \lambda = 1$. In NPF \cite{Daum2013c}, $\bm{f}^i_k\in \mathbb{R}^{4}$ is given by,
\begin{equation}
\label{PF2}
    \bm{f}^i_k = -[ -(\bm{P}^i_{k-1})^{-1}+\lambda \bm{\nabla}^2 \ln h^i_k]^{-1}(\bm{\nabla} \ln h^i_k)
\end{equation}
where $\bm{P}^i_{k-1}$ is the covariance matrix of $\bm{m}^i_{k-1}$. $\bm{\nabla}$ is the spatial vector differentiation operator $\frac{\partial}{\partial \bm{m}^i_{k-1}}$. The likelihood ${h}^{i}_k$ is given by
\begin{equation}
\label{likelihood}
{h}^{i}_k = \mathcal N(\bm{m}^i_{k|k-1}|\argmin_{\bm{z}^{r}_k } \left\| \bm{z}^r_k - \bm{m}^i_k \right\|,\bm{R})
\end{equation}
where $\bm{R}$ is the covariance matrix of the measurement noise. $\left\| \cdot \right\|$ is the $l_2$ norm. Then each particle state is updated as
\begin{equation}
\label{stateparticle}
\bm{m}^i_{k|k-1} \Leftarrow \bm{m}^i_{k|k-1} + \triangle \bm{m}^i_{k|k-1}
\end{equation}
The weights are calculated as
\begin{equation}
\label{SMCupdate}
\omega^i_{k}  = \left[ 1 - p_{D,k}+\sum _{ r = 1 }^{R_k}{ \frac { p_{D,k}h^{i}_{k} }{ \kappa_{k} +\sum _{ i = 1 }^{  N_k  }{ p_{D,k}h^{i}_{k}\omega^i_{k|k-1} }}  } \right]\omega^i_{k|k-1} 
\end{equation}
where $ {p}^i_{D,k}$ and ${\kappa}_{k}$ are the abbreviations of $ {p}^i_{D,k}( {{\bm{m} }  }^i_{ k }|  {\bm{m}}^1_{ k-1 },...,{\bm{m}}^{{N}_{k-1}}_{ k-1 })$ and ${\kappa}_{k}(\bm{Z}_k)$, respectively. $ {p}^i_{D,k}$ is the detection probability. ${\kappa}_{k}$ is the intensity function of clutter at time $k$. 
The number of targets is estimated as the sum of the weights. The states and weights of the targets $\{\tilde { \bm{m}}^j_{k},\tilde { \omega}^j_{k}\}^{\tilde { N}_k}_{j=1}$ can be calculated using e.g. K-means clustering method \cite{arthur2007k} or multi-expected a posterior (MEAP) \cite{Li2016}.

Finally, resampling is performed when the effective sample size (ESS)  \cite{kong1994sequential} is smaller than half number of particles. In the resampling step, we can obtain $\{\bm{m}^{i}_{k},\omega^{i}_{k}\}^{N_k}_{i=1}$, where  $\{\omega^{i}_{k}\}^{N_k}_{i=1} = 1/N_k$.

The NPF has mitigated the weight degeneracy problem in the SMC-PHD filter under the assumption that all targets are on the scene (visually) or active (continuously talking) during tracking. However, the LOCATA includes the practical challenges of data processing of conversational speech, such as natural speech inactivity during sentences, sporadic utterances and dialogues between multiple talkers. Therefore, the clutter intensity and detection probability should be considered for multi-speaker tracking.

\begin{algorithm}[htbp]
  \caption{NPF-SMC-PHD Filter}
  \label{alg0}
    \floatname{algorithm}{Procedure}
    \renewcommand{\algorithmicrequire}{\textbf{Input:}}
    \renewcommand{\algorithmicensure}{\textbf{Output:}}

  \begin{algorithmic}[1]
 
  \REQUIRE $\{\bm{m}^{i}_{k-1},\omega^{i}_{k-1}\}^{N_{k-1}}_{i=1}$, $\{\bm{P}^i_{k-1}\}^{N_{k-1}}_{i=1}$ and ${\bm{Z}_k}$.
  \ENSURE $\{\tilde { \bm{m}}^j_{k},\tilde { \omega}^j_{k}\}^{\tilde { N }_k}_{j=1}$,  $\{\bm{P}^i_{k}\}^{N_{k}}_{i=1}$ and $\{\bm{m}^{i}_{k},\omega^{i}_{k}\}^{N_k}_{i=1}$.
  
  \STATE {\textbf{Initialize:}$k$, $q_k$, $p_s$, $\phi_{k}$, $p_k$, $\kappa_{k}$, $P_{D,k}$, $\triangle \lambda$, $N_{ \lambda  } $, $\upsilon^i_k$, $\bm{w}^i_k$ and $N_{B}$.}
  
  \STATE \textbf{Run:} 

  \STATE  {Propagate the particle states $\{\bm{m}^{i}_{k|k-1}\}^{N_{k-1}}_{i=1}$ by Eq. \eqref{SMCprediction1}.}
  \STATE  {Calculate the particle weights $\{\omega^{i}_{k|k-1}\}^{N_{k-1}}_{i=1}$ by Eq. \eqref{SMCprediction3}.}
  \STATE  {Sample $N_{B}$ born particles  $\{\bm{m}^{i}_{k|k-1},\omega^{i}_{k|k-1}\}^{N_{k-1}+N_{B}}_{i={N_{k-1}+1}}$ uniformly around each measurement by Eq. \eqref{SMCbirth} and Eq. \eqref{SMCbirthweight}.}
  \STATE  {Combine all the particles: $\{\bm{m}^{i}_{k|k-1},\omega^{i}_{k|k-1}\}^{N_k}_{i=1} = \{\bm{m}^{i}_{k|k-1},\omega^{i}_{k|k-1}\}^{N_{k-1}}_{i=1} \cup 　\{\bm{m}^{i}_{k|k-1},\omega^{i}_{k|k-1}\}^{N_{k-1}+N_{B}}_{i={N_{k-1}+1}}$.}
  \FOR{$\lambda \in \left[ 0, \triangle \lambda, 2\triangle \lambda,\cdots , N_{ \lambda  } \triangle \lambda \right] $}
          \STATE  {Calculate the likelihood $h^i_k$ by Eq. \eqref{likelihood}.} 
          \STATE  {Calculate particle flow $\bm{f}^i_k$ by Eq. \eqref{PF2}.}  
          \STATE  {Calculate $\triangle \bm{m}^i_{k|k-1}$ by Eq. \eqref{updatem2}.}  
          \STATE  {Update each particle state by Eq. \eqref{stateparticle}.}
  \ENDFOR

  \STATE  {Update the particle weights $\{\omega^i_{k|k-1}\}^{N_k}_{i=1}$ to obtain $\{\omega^i_{k}\}^{N_k}_{i=1}$ by Eq. \eqref{SMCupdate} and calculate $\tilde { N }_k = \sum _{ i=1 }^{ { N }_{ k } }{ \omega ^{ i }_{ k } }$.   }
  \STATE  {Set $\{\bm{m}^i_{k}\}^{N_k}_{i=1}$ as $\{\bm{m}^i_{k|k-1}\}^{N_k}_{i=1}$. }
  \STATE {Cluster particles and get $\{\tilde { \bm{m}}^j_{k},\tilde { \omega}^j_{k}\}^{\tilde { N}_k}_{j=1}$ by the K-means method or MEAP}
  \STATE  {Calculate $\{\bm{P}^i_{k}\}^{N_{k}}_{i=1}$ by Kalman filter or clustered particle group.}
  \IF{ESS $< N_k/2$}
    \STATE  {Resample $\{\bm{m}^{i}_{k},\omega^{i}_{k}\}^{N_k}_{i=1}$.}       
  \ENDIF
  \end{algorithmic}
\end{algorithm}

\section{IPF-SMC-PHD FILTER}

To address the limitations of the NPF-SMC-PHD filter, we propose the IPF-SMC-PHD filter for the task 4 of the LOCATA challenge. The measurements of the IPF-SMC-PHD filter is given by the MUSIC, which is the baseline method of the LOCATA challenge. In this section, the IPF and identification of the speaker are discussed. 


\subsection{Intensity particle flow}

\begin{algorithm}[tp]
  \caption{Intensity particle flow step in the IPF-SMC-PHD filter}
  \label{alg4}
    \floatname{algorithm}{Procedure}
    \renewcommand{\algorithmicrequire}{\textbf{Input:}}
    \renewcommand{\algorithmicensure}{\textbf{Output:}}
  \begin{algorithmic}[1]
 
  \REQUIRE $\{\bm{m}^{i}_{k|k-1},\omega^{i}_{k|k-1},\bm{P}^i_{k|k-1}\}^{N_{k-1}}_{i=1}$ and ${\bm{Z}_k}$.
  \ENSURE  $\{\bm{m}^{i}_{k},\omega^{i}_{k}\}^{N_{k-1}}_{i=1}$.
  
  \STATE \textbf{Initialize:}  {$\bm{R}$, $p_{D,k}$, $\triangle \lambda$, ${\kappa}_{k}$, $\upsilon^i_k$, $\partial_{\bm{m}}$ and $\bm{w}^i_k$.}
  \FOR{$\lambda \in \left[ 0, \triangle \lambda, 2\triangle \lambda,\cdots , N_{ \lambda  } \triangle \lambda \right] $}
    \FOR{Each surviving particles}

          \STATE  { Calculate the likelihood density $\bm{h}_k^{i,r}$ based on the probability density function of the Gaussian distribution $\mathcal{N}(\bm{z}^r_k,\bm{R})$.}
          \STATE  {Calculate $\bm{\nabla} h^{i,r}_{k}$ ,$\bm{\nabla}(\bm{\nabla} h^{i,r}_{k})$ by Eq. \eqref{h1} and Eq. \eqref{h2}.}
          \STATE  {Calculate particle flow by Eq. \eqref{IPFS}.}  
          \STATE  {Calculate $\triangle \bm{m}^i_{k|k-1}$ by Eq. \eqref{updatem2}.}  
          \IF{$\triangle \bm{m}^i_{k|k-1} < \partial_{\bm{m}}$}
            \STATE  {Stop calculating the particle flow for this surviving particle. }
          \ENDIF
          \STATE  {Update each particle state by Eq. \eqref{stateparticle}.}      
          \STATE  {Update the weights of the particles $\{\omega^i_{k|k-1}\}^{N_{k-1}}_{i=1}$ to obtain $\{\omega^i_{k}\}^{N_{k-1}}_{i=1}$ by Eq. \eqref{SMCupdate}. }
    \ENDFOR
  \ENDFOR
  \STATE  {Set $\{\bm{m}^i_{k}\}^{N_{k-1}}_{i=1}$ as $\{\bm{m}^i_{k|k-1}\}^{N_{k-1}}_{i=1}$. }
  \end{algorithmic}
\end{algorithm}

The IPF is used to replace the NPF, lines 6-11 of the Algorithm 1. For decreasing the computational cost, we only update the survival particles by the IPF, since the born particles are created as the measurements. After the prediction step, the particle set is shown as $\{\bm{m}^{i}_{k|k-1},\omega^{i}_{k|k-1}\}^{N_{k-1}}_{i=1}$. Based on the intensity function \cite{vo2003sequential}, the particle flow can be calculated according to 
\begin{equation}
\begin{split}
\label{IPFS}
\bm{f}^i_k = &- [  \sum _{ r = 1 }^{R_k}{ \frac {\lambda {p}_{D,k}\bm{\nabla}(\bm{\nabla} h^{i,r}_{k}) }{G^r_k}  }  + \bm{\nabla}(\bm{\nabla}\ln({\omega}^i_{k|k-1}))]^{-1}\\
&\cdot \sum _{ r = 1 }^{R_k}{ \frac {{p}_{D,k}\bm{\nabla} h^{i,r}_{k} }{ G^r_k} }
\end{split}
\end{equation}
where
\begin{equation}
G^r_k = \kappa_{k} + \sum_{i= N_{k-1}+1}^{N_{k-1}+N_B}S^{i,r}_k + \sum _{ i = 1 }^{  N_{k-1}  }{ h^{i,r}_{k}\omega^i_{k|k-1} }
\end{equation}
\begin{equation}
\label{calculateS}
S^{i,r}_k =\gamma _k(\bm{m}^{i,r}_{k|k-1})* \max(0,1 - \sum ^{ N_{k-1} }_{ i= 1 }{ h^{i,r}_k \omega^{i,r}_{k|k-1}} )
\end{equation}
where $S^{i,r}_k$ is the birth intensity function for the $i$-th particle and the $r$-th DOA line at $k$. $\bm{\nabla}(\bm{\nabla}\ln({\omega}^i_{k|k-1}))$ is independent of the particle state and a constant for the particle flow. If we assume that likelihood model is Gaussian, the particle flow in Eq. \eqref{IPFS} may be derived analytically for particle motion. The differentiation of the likelihood $h^{i,r}_{k}$ is calculated as follows: 
\begin{equation}
\label{h1}
\bm{\nabla} h^{i,r}_{k} = - h^{i,r}_{k}\bm{R}^{-1}({\bm{f_z}\left( { \tilde{\bm{m}} }_{ k } ,{  \bm{\varsigma  }  }_{ k }\right) -\bm{z}^r_k })
\end{equation}
\begin{equation}\hspace{0in}
\label{h2}
\begin{split}
\bm{\nabla}(\bm{\nabla} h^{i,r}_{k}) = &h^{i,r}_{k}[\bm{R}^{-1}(\bm{f_z}\left( { \tilde{\bm{m}} }_{ k } ,{  \bm{\varsigma  }  }_{ k }\right) -\bm{z}^r_k )\\&(\bm{f_z}\left( { \tilde{\bm{m}} }_{ k } ,{  \bm{\varsigma  }  }_{ k }\right) -\bm{z}^r_k )^{-1}\bm{R}-\bm{R}^{-1}]
\end{split}
\end{equation}
With the increment of $\lambda$, the rate of change of $\triangle \bm{m}^i_{k|k-1}$ may decrease. If $\triangle \bm{m}^i_{k|k-1}$ is smaller than the sensor resolution $\partial_{\bm{m}}$, $\bm{m}^i_{k|k-1}$ is invariant based on Eq \eqref{stateparticle} after $\triangle \bm{m}^i_{k|k-1}$ is added to $\bm{m}^i_{k|k-1}$, which is inefficient and wasteful. So if $\triangle \bm{m}^i_{k|k-1} < \partial_{\bm{m}}$, the particle flow step would be ignored. The pseudo code of IPF in the IPF-SMC-PHD filter is shown in Algorithm \ref{alg4}. 

\begin{algorithm}[htbp]
  \caption{Identification step in the IPF-SMC-PHD filter}
  \label{alg2}
    \floatname{algorithm}{Procedure}
    \renewcommand{\algorithmicrequire}{\textbf{Input:}}
    \renewcommand{\algorithmicensure}{\textbf{Output:}}
  \begin{algorithmic}[1]
 
  \REQUIRE $\{\hat{\bm{m}}^{j}_{k-1}\}^{\hat{N}_m}_{j=1}$ and $\{\tilde{\bm{m}}^{j}_{k}\}^{\tilde{N}_{k}}_{j=1}$
  
  \ENSURE  $\{\hat{\bm{m}}^{j}_{k}\}^{\hat{N}_m}_{j=1}$ 
  
  \STATE \textbf{Initialize:}  {$\bm{d}$ and $\hat{N}_m$.}
  
  \IF{$\tilde{N}_k = \hat{N}_m$}
    \FOR{$j \in [1,..,\tilde{N}_k]$}
        \STATE  {$\hat{\bm{m}}^{j}_{k} = \underset{\tilde{\bm{m}}^{j}_k}\argmin {\left\| \hat{\bm{m}}^{j}_{k-1} - \tilde{\bm{m}}^{j}_k \right\| }$}
    \ENDFOR
  \ENDIF
  \IF{$\tilde{N}_k > \hat{N}_m$}
    \FOR{$j \in [1,..,\hat{N}_m]$}
        \STATE  {$\hat{\bm{m}}^{j}_{k} = \underset{\tilde{\bm{m}}^{j}_k}\argmin {\left\| \hat{\bm{m}}^{j}_{k-1} - \tilde{\bm{m}}^{j}_k \right\| }$}
    \ENDFOR
  \ENDIF
  \IF{$\tilde{N}_k < \hat{N}_m$}
    \FOR{$j \in [1,..,\tilde{N}_k]$}
        \STATE  {$\hat{\bm{m}}^{j}_{k} = \underset{\tilde{\bm{m}}^{j}_k}\argmin {\left\| \hat{\bm{m}}^{j}_{k-1} - \tilde{\bm{m}}^{j}_k \right\| }$}
        \IF{$\left\| \hat{\bm{m}}^{j}_{k-1} -\hat{\bm{m}}^{j}_{k}  \right\| > \bm{d}$}
            \STATE  {$\hat{\bm{m}}^{j}_{k} = { \bm{f} }_{ \tilde { { \bm{m} } }  }\left( \hat{\bm{m}}^{j}_{k-1} ,{ \bm{ \tau  } }_{ k } \right)$ }
        \ENDIF
    \ENDFOR
  \ENDIF
  \end{algorithmic}
\end{algorithm}

\subsection{Identification of the speaker}
As all estimated positions must be associated with an identity (ID) in the LOCATA challenge, the estimates resulting from the IPF-SMC-PHD filter should consider false tracks, missing tracks, broken tracks and track swaps. However, the PHD filter does not consider the identity of speakers. An assistant identifier should be added. Since the number of speakers is not known, the identification problem is normally solved by the Blind Source Separation (BSS) method. However, the BSS has a high computational complexity. As the IPF-SMC-PHD filter can provide the estimate of the speaker state, in our proposed method, the speaker identity is estimated by the speaker states under the assumption that it is not changing abruptly in subsequent frames. Although the number of speakers at each frame $\{\tilde{N}_k\}^k_{i=1}$ has been estimated at the line 12 of the Algorithm 1, the estimated number is varying due to the noise and undetected DOA lines. For smoothing the trajectory of speakers, we assume the mean number of speakers $\hat{N}_m$ is given by:
\begin{equation}
\hat{N}_m = \frac{\sum^{k}_{i=1}{N_k}}{k} 
\end{equation}

For each frame, if the number of the estimated speakers $N_k$ is larger than $\hat{N}_m$ at frame $k$, it may imply that the noises are estimated as the speakers. To detect the noise, the distance from the estimated speaker state at $k$ and the speaker state $k-1$ is considered. As we assume the positions are not changed abruptly in subsequent frames, the estimated speaker state at $k$ with less distance to the state at $k-1$ is considered as the speaker state at $k$, where $j\in {1,...,\hat{N}_m}$. If the number of the estimated speakers $N_k$ is less than $\hat{N}_m$ at frame $k$, it may imply the miss detection of speakers. The undetected speaker states are updated by the velocity as Eq. \eqref{trans}. If the number of the estimated speakers $N_k$ is equal to $\hat{N}_m$ at frame $k$. The identify of the speaker is given based on the distance from the estimated speaker state to the speaker state at last frame. The pseudo code of identification step in the IPF-SMC-PHD filter is shown in Algorithm \ref{alg2}, where $\{\hat{\bm{m}}^{j}_{k-1}\}^{\hat{N}_m}_{j=1}$ is the set of the speaker states which is ordered by ID, for example, $\hat{\bm{m}}^{1}_{k-1}$ means the state of the first speaker.

\section{Experimental Results}

In this section, the proposed algorithm is compared with its baseline counterparts including the NPF-SMC-PHD \cite{yang2018particle}, SMC-PHD algorithms \cite{kilic2016mean} and the baseline MUSIC of the LOCATA dataset \cite{lollmannlocata}. The parameters of the PHD filter and particle flow filters are set as in \cite{kilic2016mean} and \cite{yang2017audiovisual}. The number of particles per speaker is 50 and the particles are spread randomly in the tracking area. The experiments are run in Matlab on Windows 7 with Intel i7 (3.2 GHz).

The LOCATA dataset consists of sequences where multiple speakers may speak or walk. Those actions are recorded by four circular eight-element microphone arrays at 48 kHz. Although the baseline MUSIC method is provided by the LOCATA challenge, the MUSIC only considers one speaker. So we consider more signal subspaces to calculate the DOA lines than the baseline MUSIC. The parameters of the microphone arrays are shown in Table 1, which are chosen based on the ground truth dataset of the Task 1 and Task 2.

\begin{table}
\centering
\caption{The index of used microphones and the number of the subspaces for the DICIT array, Eigenmike array, Robot head and Hearing aids.}
\begin{tabular}{ |c|c|c| } 
\hline
  Array &  Index  & Number \\
\hline
DICIT & 5,6,7,9,10 & 1,2\\
\hline
Eigenmike &  1,...,32 & 1,2,3,4,5\\
\hline
Robot head&  1,...,12 & 1,2,3,4\\
\hline
Hearing aids &  1,2,3,4 & 3,4\\
\hline
\end{tabular}
\label{table:compareall}
\end{table}

\begin{figure}[h]
    \centering
     \includegraphics[width=\linewidth]{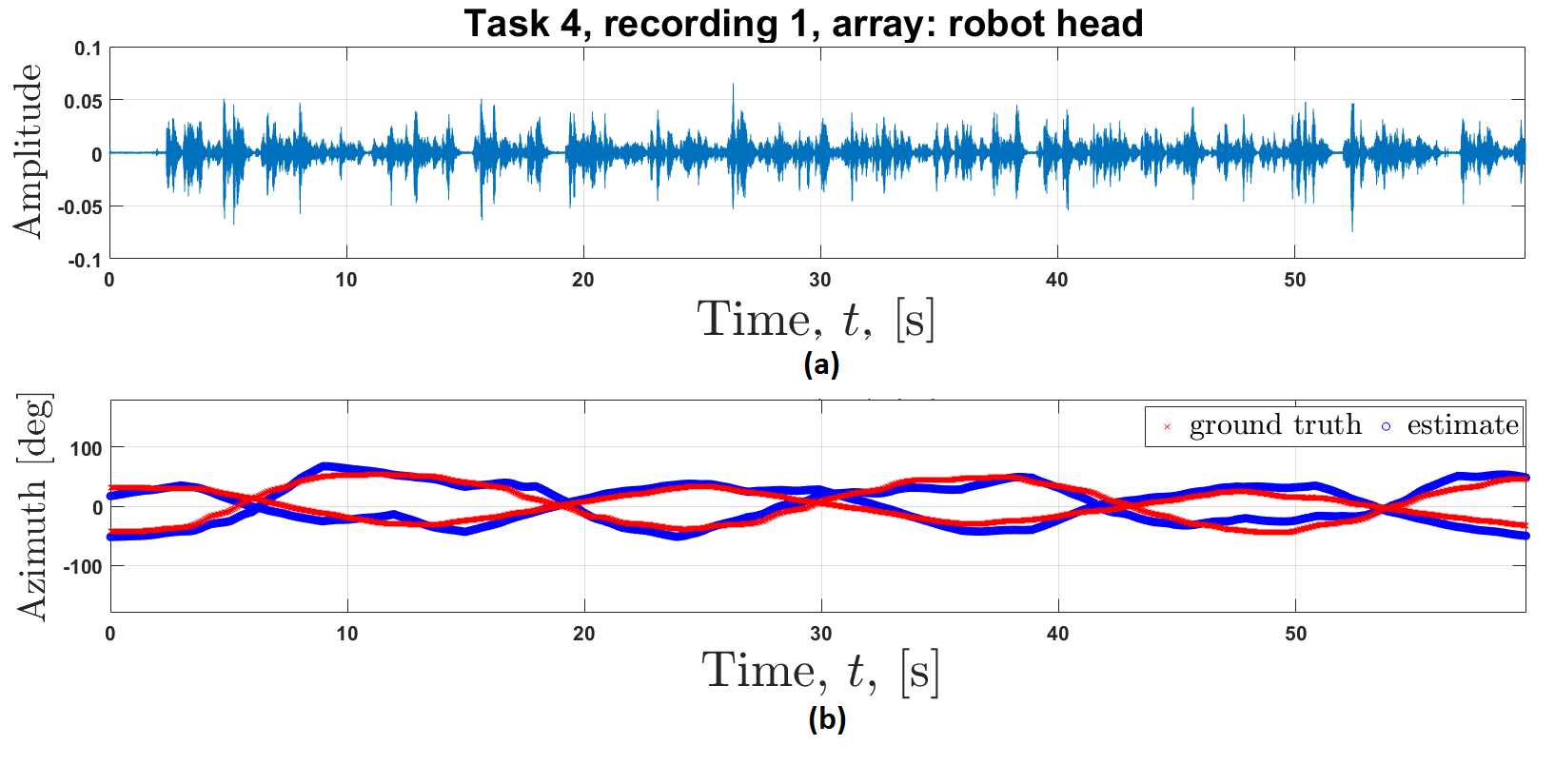}
    \caption{The audio signal is illustrated in (a), and (b) shows the speaker state estimated by the IPF-SMC-PHD filter and ground truth speaker state with the Robot head array on recording 1 of the evaluation data of the task 4.}
    \label{compare5}
\end{figure}

\begin{table}
\centering
\caption{The OSPA for the IPF-SMC-PHD, NPF-SMC-PHD, ZPF-SMC-PHD filters and MUSIC algorithm in terms of the OSPA error on the Locata task 4.}
\hspace*{0cm}
\setlength{\tabcolsep}{4.5pt}
\renewcommand{\arraystretch}{1.5}
\begin{tabular}{ |c|c|c|c|c|c|c|c|c|c| } 
\hline
 Array & Recording & IPF & NPF  &  SMC & MUSIC\\
\hline
           & 1 & $\bm{1.084}$ & 1.178  & 1.247 & 1.875 \\\cline{2-2}
Robot head & 2 & $\bm{1.079}$ & 1.165  & 1.242 & 1.753 \\\cline{2-2}
           & 3 & $\bm{1.093}$ & 1.205 & 1.253 & 1.897 \\
\hline
           & 1 & $\bm{4.826}$ & 5.893  & 7.089 & 10.357 \\\cline{2-2}
DICIT      & 2 & $\bm{4.543}$ & 5.407   & 6.580 & 10.182\\\cline{2-2}
           & 3 & $\bm{5.405}$ & 6.777   & 7.860  & 11.057\\
\hline 
           & 1 & $\bm{4.833}$ & 5.894  & 7.091 & 10.360\\\cline{2-2}
Hearing aids      & 2 & $\bm{4.591}$ & 5.603  & 6.736 & 9.848\\\cline{2-2}
           & 3 & $\bm{5.310}$  & 6.507  & 7.895 & 11.490\\
\hline
           & 1 & $\bm{1.465}$ & 1.559  & 1.568 & 2.288\\\cline{2-2}
Eigenmike  & 2 & $\bm{1.295}$ & 1.461  & 1.616 & 2.212\\\cline{2-2}
           & 3 & $\bm{1.399}$ & 1.503  & 1.656 & 2.429\\
\hline
   \multicolumn{2}{| c |}{\textbf{Average OSPA}}&$\bm{3.077}$ &$3.679$   & $4.319$ & $6.312$ \\
\hline
\end{tabular}

\label{table:compareal2}
\end{table}
 
Due to the space limitation, we only show the tracking result on recording 1 on the robot head. Figure 1a shows the signal representation of recording 1 of task 4. Speaker states are indicated with blue and red line in Figure 1b, respectively for the IPF-SMC-PHD and ground truth. Here, we performed down-sampling to the plots for better visualization. At the beginning of the recording, the speakers are silent and the estimates are calculated when the speakers start to talk. Although the filter can detect the occlusions, the error increases when the occlusions happens.

The Optimal Sub-pattern Assignment (OSPA) for trackers  \cite{ristic2011metric}, which gives a combined score for the estimation performance in the number of sources and their positions, is used to evaluate the tracking accuracy. 
Table \ref{table:compareal2} reports the average OSPA over 10 random tests. With the contribution of the IPF, 16\% reduction in tracking error has been achieved as compared with the NPF-SMC-PHD filter. In addition, the IPF-SMC-PHD filter also improves the estimation accuracy by 29\% and 51\% over the SMC-PHD and baseline MUSIC method, respectively. However, the running time of IPF (about 10s/frame) is three times and ten times of the SMC-PHD filter (about 3s/frame) and the MUSIC method (about 1s/frame), respectively.




\section{Conclusion}

We have presented a novel IPF-SMC-PHD filter for audio multi-speaker tracking by smoothly migrating the particles. The proposed algorithm has been tested on the task 4 of the LOCATA dataset. The experimental results show that the proposed filter offers a higher tracking accuracy than the baseline methods with a higher computational cost.

\section{Acknowledgment}
\label{sec:ack}

This work was supported by the EPSRC Programme Grant S3A: Future Spatial Audio for an Immersive Listener Experience at Home (EP/L000539/1), the BBC as part of the BBC Audio Research Partnership, the China Scholarship Council (CSC), and in part by the EPSRC grant EP/K014307/2.

\bibliographystyle{IEEEtran}
\bibliography{refs18}
%
%
%
%
%
%
%
%
%

\end{sloppy}
\end{document}